\documentstyle[12pt]{article}

\newcommand{ \bb }{ $2\nu\beta\beta \ $ }

\begin{document}

\begin{center}
APPROXIMATE SYMMETRIES IN NUCLEI AND THE \bb - DECAY RATE
\end{center}

\centerline {\it {O.A.Rumyantsev, M.H.Urin}}
\medskip
\centerline {\it Moscow Engineering-Physics Institute, 115409 Moscow, Russia}

\bigskip
A nonstandard method for calculating the nuclear $2\nu\beta\beta$ -
decay amplitude is proposed. The method is based on the explicit use of
those approximate symmetries of a nuclear hamiltonian, which correspond
to the operators of allowed $\beta$ -- transitions. Within the
framework of the proposed method the mentioned amplitude is calculated
for a wide range of nuclei. The model parameters used in calculations
are taken from independent data.  Calculated \bb half-lifes are
compared with known experimental data.
\bigskip

1. An analysis of a great body of known experimental data (see e.g.
ref. \cite{Mor94}) allows one to conclude that the
\bb - decay rate is hindered as compared with the rates
evaluated within the independent-quasiparticle approximation.  The
hindrance is caused by existence of the isobaric analogue and
Gamow-Teller nuclear collective states (IAS and GTS) with relatively
large excitation energy. These states each exhaust the main part of the
Fermi and Gamow-Teller strength, respectively.  To describe the
hindrance the different versions of the quasiparticle random phase
approximation (QRPA) are used for calculations of the nuclear \bb -
decay amplitude. It has been found (see e.g. refs.
\cite{Vog88,Mut95,Civ95}) that the
calculated amplitudes are very sensitive to the particle-particle
interaction strength. For this reason the predictable power of the QRPA
current versions is poor. As reported recently, the similar instability
of the calculated $0\nu\beta\beta$ - decay amplitude has been also
found \cite{Hir95,Pan96}.  We believe that the mentioned difficulty
caused by inconsistency of the QRPA current versions can be presently
overcome by the explicit use of the approximate isospin and
spin-isospin symmetries of a nuclear hamiltonian. Early attempts along
this line have been undertaken in refs. \cite{Ber90,Rum95}.

It is well-known that the Fermi-type excitations can be correctly
described only with consideration for the isospin symmetry of a nuclear
hamiltonian. Such a consideration allows one to reproduce in model
calculations the experimental fact, that IAS exhausts more than 95\%
the Fermi sum rule $N-Z$, as well as to evaluate correctly the small
Fermi-strength of the low-energy IAS satellites populated virtually in
the $2\nu\beta\beta$ - decay process. If the isospin symmetry were
exact, the IAS would exhaust 100\% of the Fermi strength and the Fermi
nuclear amplitude for the $2\nu\beta\beta$ - decay to the ground or
low-excited states of a product nucleus would be equal to zero. The
above statements allow one to describe the Fermi-type excitations
within the framework of the perturbation theory with respect to the
variable part of the nuclear mean Coulomb field. This latter is the
main source of the isospin-symmetry violation in intermediate and heavy
mass nuclei (see e.g. ref.
\cite{Rum93} and refs. therein).

The similar situation takes place at the description Gamow-Teller (GT)
excitations. It is experimentally established that in intermediate and
heavy mass nuclei the GTS exhausts about 70\% - 80\% sum rule $3e_q^2 (
N-Z ) $, where $e_q \simeq 0.8$ is ''a effective charge'' describing,
in particular, the renormalization of the axial constant of the weak
interaction in nuclei. The mentioned experimental fact can be
considered as an evidence of approximate conservation of the
appropriate spin-isospin symmetry of a nuclear hamiltonian (see in this
connection ref. \cite{Gap92}).  This symmetry should be taken into
consideration for correct description of both the GTS low-energy
satellites and the nuclear
\bb - decay amplitude. If the mentioned symmetry were exact,
the amplitude would be equal to zero. Because the \bb-decay hindrance
is connected with formation of the collective GTS, the hindrance can be
also considered as a result of partial conservation of the appropriate
spin-isospin symmetry.

In the presented work a method for calculating the GT nuclear
$2\nu\beta\beta$ - decay amplitude is proposed. The method is based on
the explicit consideration for approximate conservation of the
symmetries corresponding to operators of the allowed Fermi and GT
$\beta$ - transitions. The $2\nu\beta^-\beta^-$ and $2\nu\beta^+
\beta^+$ (including the electron capture) half-lifes for a wide range
of nuclei are calculated within the framework of the simplest version
of the proposed method. The model parameters used in calculation are
taken from independent data. Calculation results are compared with
known experimental data.

2. To clarify the main point of the proposed method we derive the basic
formula for nuclear \bb - decay amplitude by the same way for both the
Fermi and GT amplitudes. At first we consider the $2\nu\beta^-\beta^-$
- decay. Let $\vert i \rangle$ and $\vert f
\rangle$ be the wave functions of the ground state, respectively,
of a double-even parent-nucleus $(N+2,Z)$ with energy $E_i$ and of a
product-nucleus $(N,Z+2)$ with the energy $E_f$, $G^{(\pm)}$ be the
operators of allowed Fermi ($G^{(\pm)} =\sum_a
\tau^{(\pm)} (a)$) or Gamow-Teller ($G^{(\pm)} = \sum_a\vec
\sigma (a) \tau^{(\pm)}(a)$) $\beta$ - transitions.

We start with the zero approximation, in which symmetries corresponding
to operators $G^{(\pm)}$ are exact. In this approximation the states
$\vert G \rangle$ (IAS and GTS) of an isobaric nucleus $(N + 1,Z + 1)$
: (i) each exhaust 100\% of the corresponding sum rule, so that

\begin{equation}
\vert G \rangle = k_G \ G^{( - )} \vert i \rangle,\ \ G^{( + )} \vert i
\rangle = G^{( + )} \vert f \rangle=0; \
k_G = \langle i \vert [G^{(+)},G{(-)}] \vert i \rangle^{-1/2}\ ;
\end{equation}

\noindent (ii) have the same energy $E_G$; (iii) are orthogonal to
other states $\vert A \rangle$ of the same isobaric nucleus:

\begin{equation}
\langle G \vert A \rangle = 0;
\ \ \ G^{(+)} \vert A \rangle =0\ .
\end{equation}

\noindent For the further analysis of the nuclear \bb - decay amplitudes
it is necessary to consider also the following excited states of a
product-nucleus:

\begin{eqnarray}
&\vert GA \rangle = k_A \ G^{(-)} \vert A \rangle,\ \ E_{GA} = E_G -
E_i + E_A, \ \ k_A = \langle A \vert [ G^{(+)}, G^{(-)}]
\vert A \rangle^{-1/2}\ , \\ \nonumber \\
&\vert DG \rangle = k_D \ G^{(-)} G^{(-)} \vert i
\rangle,\ \ E_{DG} = E_G - E_i + E_G,\ \ k_D = \langle i \vert
G^{(+)} G^{(+)} G^{(-)} G^{(-)} \vert i
\rangle^{-1/2}\ .
\end{eqnarray}

\noindent Low-energy states $\vert A \rangle$ are called the
anti-IAS or anti-GTS, whereas states $\vert DG \rangle$ are called the
double-IAS or double-GTS.

Let $U$ be the nuclear mean field, so that $U = \sum_a U({\bf r}_a)$
and $[U,G^{(-)}]=U_G^{(-)} \ne 0$. The part of the nuclear mean field
$U_G = \sum_a U_G({\bf r}_a)$, which leads to nonzero operator
$U_G^{(-)}$, is the main source of violating the symmetry connected
with operator $G^{(-)}$. Other sources are less essential and briefly
considered below. In the lowest order in the $U_G$ the energy of state
(1) is equal to

\begin{equation}
E_G - E_i = k_G^2 \langle i \vert [G^{(+)}, U_G^{(-)}] \vert i \rangle
\equiv \Delta_G\ .
\end{equation}

Let us turn to the expression for the nuclear $2\nu\beta^-\beta^-$ -
decay amplitude (see e.g. refs. \cite{Doi88,Doi92}):

\begin{equation}
\label{Mbb}
M_G = \sum_S \langle f \vert G^{(-)} \vert S \rangle \langle S \vert
G^{(-)} \vert i \rangle \omega_S^{-1}\ ,
\end{equation}

\noindent where the sum is taken over intermediate states of an
isobaric nucleus $(N + 1,Z + 1)$; $\omega_S = E_S - (E_i + E_f)/2$ is
the excitation energy of the intermediate state. The $2\nu\beta^+
\beta^+$ - decay amplitude is equal to conjugate
value $M_G^* $. In the zero approximation the \bb - decay is doubly
forbidden $(M_G = 0)$, as it follows from eqs.  (1)-(6) after replacing
$\vert S \rangle \to \vert A \rangle,
\vert G \rangle$. Mixing of nuclear states due to the single-particle
field $U_G$ results in nonzero value of amplitude $M_G$. We keep those
corrections to wave functions of isobaric nuclei, which provide the
contribution to the amplitude $M_G$ of the same order perturbation with
respect to $U_G$:

\begin{equation}
\label{cor0}
\begin{array}{c}
\vert G \rangle \to \vert G \rangle + {\displaystyle \sum_A}
\alpha_{A,i} \vert A \rangle;
\ \ \ \vert A \rangle \to \vert A \rangle + \alpha_{i, A} \vert G
\rangle\ ; \ \ \ \alpha_{A,i}=-\alpha_{i,A}^*;\\ \\
\vert f \rangle \to \vert f \rangle + {\displaystyle \sum_A}
\alpha_{A, f} \vert GA \rangle + \beta_{i, f} \vert DG \rangle\ .
\end{array}
\end{equation}

\noindent With the help of eqs. (1)-(4) the first-order mixing
amplitudes $\alpha$ and the second-order amplitude $\beta$ can be
expressed in terms of matrix elements of the single-particle
charge-exchange field $U_G^{(-)}$:

\begin{equation}
\label{cor1}
\begin{array}{c}
\alpha_{A,i} = \frac{\displaystyle \langle A
\vert U \vert G \rangle}{\displaystyle E_G - E_i} = k_G
\frac{\displaystyle \langle A \vert U_G^{(-)} \vert i
\rangle}{\displaystyle \omega_G - \omega_A}, \ \ \
\alpha^*_{A, f} = \frac{\displaystyle \langle f \vert U \vert GA
\rangle}{\displaystyle E_f - E_{GA}}
= - k_A \frac{\displaystyle \langle f \vert U_G^{(-)} \vert A
\rangle}{\displaystyle \omega_G + \omega_A},\\ \\
\beta^*_{i, f} = {\displaystyle \sum_A} \frac{\displaystyle \langle f
\vert U \vert GA \rangle \langle GA \vert U \vert DG \rangle}{\displaystyle
(E_{GA} - E_f)(E_{DG} - E_f)} = k_D {\displaystyle \sum_A}
\frac{\displaystyle \langle f \vert U_G^{(-)}
\vert A \rangle \langle A \vert U_G^{(-)} \vert i \rangle}{\displaystyle
(\omega_G +\omega_A ) \omega_G}\ .
\end{array}
\end{equation}

\noindent After substitution of the eqs. (\ref{cor0}),(\ref{cor1}) in eq.
(\ref{Mbb}) with the use of eqs. (1)--(4) we get:

$$ M_G = \omega_G^{-2} \sum_A \langle f \vert U_G^{( - )} \vert A
\rangle
\langle A \vert U_G^{( - )} \vert i \rangle \omega_A^{-1}\ .
$$

\noindent This equation can be presented in the form, which
is similar to the starting expression (6):

\begin{equation}
\label{MG}
M_G = \omega_G^{-2} \sum_S \langle f \vert V_G^{( - )}
\vert S \rangle \langle S \vert V_G^{( - )} \vert i \rangle
\omega_S^{-1}\ ,
\end{equation}

\noindent where $V_G^{(-)} \equiv U_G^{(-)} - \Delta_G G^{(-)}, \
\Delta_G$ is determined by eq. (5). Indeed, the equality
$\langle G \vert V_G^{(-)} \vert i \rangle = 0$ and eqs. (1),(2) allow
one to include formally the state $\vert G \rangle$ in sum (\ref{MG}).

The fundamental character of formula (\ref{MG}) is explained by the
following. (i) In this formula the approximate symmetries of a nuclear
hamiltonian are explicitly taken into account. (ii) The hindrance of
the strength of intermediate $\beta$ -- transitions due to formation of
the collective states (IAS and GTS) is correctly taken into
consideration in rather general form. (iii) Inasmuch as the main part
of polarization effects caused by a quasiparticle interaction has been
taken into account, rather simple nuclear models can be used for
calculating the intermediate-state wave functions and energies in eq.
(\ref{MG}).

3. As the first step in evaluation of the \bb - decay amplitude
according to eq. (\ref{MG}) we use: (i) realistic phenomenological mean
nuclear field with consideration for partial self-consistency
conditions; (ii) the BCS model for nuclei with strong nucleon pairing;
(iii) the pair-vibration model for description of the ''magic $\pm$ two
nucleon'' subsystem.

We use the following phenomenological quantities involved in the model
hamiltonian. The isoscalar part of the nuclear mean field $U_0 ({\bf
r})$ is the sum of central and spin-orbital parts:

\begin{equation}
U_0 ({\bf r}) =U_0(r) + U_{SO}({\bf r});\ \ \ U_{SO}({\bf r}) =
U_{SO}(r)({\bf l s})\ .
\end{equation}

\noindent As an isovector part of the particle-hole interaction
the Landau-Migdal forces are used:

\begin{equation}
\label{LM}
{\cal F} ( {\bf r}_1, {\bf r}_2) = ( F' + G'\vec \sigma_1 \vec
\sigma_2 ) \vec \tau_1 \vec \tau_2 \delta ( {\bf r}_1 - {\bf r}_2)\ ,
\end{equation}

\noindent where $F'$ and $G'$ are phenomenological constants.
Other parts of the nuclear mean field are calculated by the
self-consistent way. As a result of the isospin symmetry of the nuclear
hamiltonian, the self-consistency condition, which relates the
isovector part of nuclear mean field $U_{SY} = {1 \over 2} \tau^{(3)} v
(r) $ and neutron-excess density $\rho(r) = \rho^n(r) - \rho^p(r)$ via
the interaction (\ref{LM}), takes place (see e.g. ref. \cite{Bir74}):
$v(r)=2F'\rho(r)$.  The nuclear mean Coulomb field $U_C(r)$ is
calculated within the Hartree approximation via the proton density
$\rho^p$.

As mentioned above, the mean Coulomb field $U_C = \sum_a U_C(r_a)
(1-\tau^{(3)}(a))/2$ is the main source of the isospin-symmetry
violation in intermediate and heavy mass nuclei. In particular, this
field results in the shift of the IAS energy from energy $E_i$.
According to eq. (5) the shift equals the Coulomb displacement energy
$\Delta_C$:

\begin{equation}
E_{IAS} - E_i = \Delta_C = (N+2-Z)^{-1} \int U_C(r) \rho(r)d{\bf r}.
\end{equation}

\noindent Here $\rho(r) $ is the above-mentioned neutron-excess density
in the ground state of the parent nucleus:

\begin{equation}
\rho (r) = {1\over 4\pi} \left ( \sum_\nu R^2_\nu(r)(2j_\nu + 1)n_\nu -
\sum_\pi R^2_\pi (r) (2j_\pi + 1) n_\pi\right ),
\end{equation}

\noindent where $R_\lambda (r)$ are the radial single-neutron
($\lambda = \nu$) and single-proton ($\lambda = \pi$) wave functions;
$\lambda = n_r, l, j$ is the set of the single-particle quantum
numbers; $n_\lambda$ are occupation numbers satisfying the equations:

\begin{equation}
\sum_\nu (2j_\nu + 1)n_\nu = N + 2,\ \ \ \sum_\pi (2j_\pi + 1)n_\pi =Z.
\end{equation}

\noindent For nuclei with strong pairing in any nucleon subsystem
the occupation numbers are $n_{\lambda} = v^2_\lambda = 1 -
u^2_\lambda$, where $v_\lambda$ and $u_\lambda$ are the
Bogoliubov-transformation coefficients. In the case that a nucleon
subsystem is ''magic $\pm$ two nucleons'' one the occupation factors
are: $n_\lambda = n^m_\lambda + c^2_\lambda(1-2 n^m_\lambda)$, where
$n^m_\lambda$ are the occupations numbers for the magic subsystem;
$c_\lambda$ are the coefficients determining the pair-vibration wave
function and satisfying the normalization condition: $\sum_\lambda
c^2_\lambda (1-2n^m_\lambda )(2j_\lambda + 1) = \pm 2$.

In view of a high degree of the isospin-symmetry conservation in nuclei
we omit the analysis of the relatively small Fermi-\bb -- decay
amplitude and turn to the GT amplitude $M_{GT}$.  The mean Coulomb
field results in the same (in the lowest order of the perturbation
theory) energy shift $\Delta_C$ (12) for the both IAS and GTS. However,
the main source of violation of the spin-isospin symmetry corresponding
to operator $G^{(-)} =
\sum_a \vec \sigma_a \tau^{(-)}_a$ is the spin-orbital part of
nuclear mean field (10): $U_{SO} =\sum_a U_{SO}({\bf r}_a)$.  In
particular, this field results in the energy shift $\Delta_{SO}$ of the
GTS from the IAS energy. According to eq. (5) we have:

\begin{equation}
\begin{array}{c}
\Delta_{SO} = \frac{\displaystyle 1}{\displaystyle 3(N+2-Z)}
\langle i \vert [G^{(+)} [U_{SO},G^{(-)}]] \vert i \rangle =
-\frac{\displaystyle 4}{\displaystyle 3(N+2-Z)}
\langle i \vert U_{SO} \vert i \rangle;\\ \\
\langle i \vert U_{SO} \vert i \rangle =
{\displaystyle \sum_{\lambda=\pi,\nu}} \langle \lambda \vert U_{SO}(r)
\vert \lambda \rangle ({\bf l s})_{j_\lambda l_\lambda} (2j_\lambda+1)
n_\lambda\ .
\end{array}
\end{equation}

\noindent Here $\langle \pi \vert U_{SO} (r) \vert \nu \rangle$ is the
radial matrix element; $n_\lambda$ are occupation numbers satisfying
eqs. (14).

In the case of GT excitations the mixing (spin-orbital) field in eq.
(9) has the form: $V_G^{(-)} = \sum_a V_{SO}(a) \tau^{(-)}_a$, where
$V_{SO}(a)=U_{SO}(r_a)[{\bf ls},\vec \sigma] - \Delta_{SO} \vec
\sigma$. Calculation of amplitude $M_{GT}$ according to eq. (\ref{MG})
within the framework of the BCS model results in the expression:

\begin{equation}
\label{MGT}
M_{GT} =e^2_q \omega_{GTS}^{-2} {\displaystyle \sum_{\pi,\nu}}
\Biggl( (2l_\pi+1)(j_\pi-j_\nu) \langle \pi \vert U_{SO}(r) \vert \nu
\rangle - \Delta_{SO} \langle \pi \vert \nu \rangle \Biggr)^2 \langle \pi
\Vert \sigma \Vert \nu \rangle^2 u_\pi v_\pi u_\nu v_\nu
\omega_{\pi \nu}^{-1}\ .
\end{equation}

\noindent Here $\langle \pi \vert \nu \rangle = {\displaystyle  \int}
R_\pi(r) R_\nu(r) r^2 dr$ is the overlap integral; $\langle \pi \Vert
\sigma \Vert \nu \rangle$ is the reduced matrix element;
$\omega_{\pi\nu} = {\cal E}_\pi + {\cal E}_\nu$ is the excitation
energy of the two-quasiparticle state, $ {\cal E}_\lambda =
\sqrt{(\epsilon_\lambda - \mu )^2 + \Delta^2}$ is the
single-quasiparticle energy for subsystem with strong nucleon pairing,
$\epsilon_\lambda$ is the energy of single-particle level, $\Delta$ is
the energy gap. If nucleon subsystem is magic in final (initial) state
and ''magic $\pm$ two nucleon'' in initial (final) state the following
replacements in eq. (16) should be made: $u_\lambda v_\lambda \to
c_\lambda$ and ${\cal E}_\lambda =
\vert \Delta + \epsilon_\lambda - \epsilon_1 \vert$ for 
particle pair-vibrations or ${\cal E}_\lambda=\vert \Delta + \epsilon_0
- \epsilon_\lambda \vert$ for hole pair-vibrations, where $\epsilon_1$
($\epsilon_0$) is the energy of the first empty (last filled)
single-particle level in the magic subsystem; $2\Delta$ is the
pair-vibration state energy, which coincides with the pair energy $P$
for the subsystem ''magic $\pm$ two nucleons''. The hindrance factor
$h_{GT}$, which can be considered as a perturbation theory parameter,
is estimated as follows: $h_{GT} = M_{GT}/M_{GT}^0$, where amplitude
$M_{GT}^0$ is evaluated accordind to eq. (\ref{Mbb}) within the
framework of the BCS model without consideration for the spin-isospin
symmetry:

\begin{equation}
\label{MGT0}
M_{GT}^0 = e^2_q \sum_{\pi,\nu} \langle \pi \vert \nu \rangle^2 \langle
\pi \Vert \sigma \Vert \nu \rangle^2 u_\pi v_\nu u_\nu v_\nu
\omega_{\pi\nu}^{-1}\ .
\end{equation}

In the conclusion of this section we briefly consider other sources of
the spin-isospin symmetry violation of the nuclear hamiltonian.  The
inequality $F'\ne G'$ in eq. (\ref{LM}) results in appearance of the
mixing field $V_\delta^{(-)} = \delta \sum_a v(r_a)
\tau^{(-)}(a)$, where $\delta=( G'-F')/F'$, $v(r)$ is the symmetry
potential considered previously. In view of smooth radial dependence of
$v(r)$ (as well as in the case of mean Coulomb field $U_C$) the
contribution of field $V_\delta^{(-)}$ to the GTS energy should be only
taken into account \cite{Gap92}. By analogy to eq.  (12) we have:

$$
\Delta_\delta = \delta (N+2-Z)^{-1} \int v(r) \rho(r) d{\bf r} \ .
$$

\noindent Thus, the GTS excitation energy in eq. (\ref{MGT}) equals
$\omega_{GTS} = \Delta_C + \Delta_{SO} + \Delta_\delta + {1\over 2} (
E_i - E_f ) $.

Another source of the spin-isospin symmetry violation is the
particle-particle interaction and, in particular, the nucleon pairing.
The relative contribution of this source to the violation intensity can
be estimated as the ratio $\Delta_{n,p}/\Delta_{ls}$ ($\Delta_{ls}$ is
the mean energy of the spin-orbit splitting near the Fermi energy),
which is rather small ($\le 20\%$) for intermediate and heavy mass
nuclei.  Thus, we do not expect of the essential dependence of
amplitude $M_{GT}$ calculated within the framework of the proposed
method on the particle-particle interaction strength in contrast to the
standard methods (see e.g. \cite{Vog88}).

4. Let us turn to the choice of model parameters and to calculation
results. The parametrization and phenomenological parameters of the
isoscalar part of the nuclear mean field are given in detail in ref.
\cite{Che67}. The intensity $F'= 300\ MeV\ fm^3$ in eq. (\ref{LM}) is chosen
to describe the experimental neutron and proton bound-energy difference
for nuclei with rather large neutron excess $^{48}Ca,^{68}Ni, ^{132}Sn,
^{208}Pb$ \cite{Wap85}, for which the difference is mainly determined
by the symmetry potential and mean Coulomb field. The intensity $G'=
255\ MeV\ fm^3$ in eq. (\ref{LM}) is chosen according to ref.
\cite{Mig83}.
The parameters $\mu_{n,p}$ and $\Delta_{n,p}$, which are used for
calculation of the Bogoliubov-transformation coefficients, were found
for each subsystem with strong nucleon pairing so that to satisfy eqs.
(14) and to describe the experimental pairing energies $P = 2
\min_\lambda {\cal E}_\lambda$ taken according to ref. \cite{Wap85}.
The pair-vibration state energies $2\Delta=P$ are calculated on the
experimental pairing energies \cite{Wap85}. Coefficients $c_\lambda$
determining the pair-vibration state wave function for a nucleon
subsystem ''magic + or -- two nucleons'', are calculated according to
equations $c_\lambda\sim(1-2n_\lambda)/(\epsilon_\lambda -
\epsilon_1 + 2\Delta ) $ or $c_\lambda\sim (1-2n_\lambda)/(\epsilon_0 -
\epsilon_\lambda + 2\Delta)$ with taking into account the normalization
conditions given in Sect.3 \cite{Sap85}. For all considered nuclei the
single-particle basis including all bound states as well as the
quasi-bound states up to $\sim 5 MeV$ is used. If the nucleon subsystem
in a parent nucleus is ''magic $\pm$ two nucleons'', and in the
product-nucleus is ''magic $\pm$ four nucleons'', the calculations were
performed within the framework of the BCS model with the use of
$u_\lambda, v_\lambda$ factors calculated for a ''magic $\pm$ four
nucleons'' subsystem.

Calculation results and corresponding experimental data for a wide
range of nuclei are given in Table 1. The $M_{GT}$ amplitudes
calculated by eq. (\ref{MGT}) are given (in $m_e^{-1}$, $m_e$ is the
electron mass) in column 5. The $M^0_{GT}$ amplitudes were calculated
by eq. (\ref{MGT0}) with the use of the same model parameters and the
same number of basic states. The calculated hindrance factors $h_{GT}$
are given in column 6. Half-lifes $T_{1/2}^{calc}$ calculated by
formula $(T_{1/2})^{-1}=G_{2\nu}
\vert M_{GT}\vert^2$ are given in column 8. The lepton factors
$G_{2\nu} $ taken from refs. \cite{Mor94,Doi88,Doi92} are also given
(column 3). We calculated also the $2\nu\beta^+\beta^+$ half-lifes
(including the electron capture) for those nuclei, for which appearance
of relevant experimental data is expected.

Two main conclusions follow from the data given in Table 1.  (i) The
calculated hindrance factors $h_{GT}$ are found to be sufficiently
small to justify the use of the perturbation theory with respect to the
spin-orbital part of nuclear mean field.  (ii) Except for the $Te$
isotopes, the calculated amplitudes $M_{GT}$ within the factor 2 -- 3
are in agreement with corresponding amplitudes $M^{exp}_{GT}$ deduced
from experimental data.

The following step in the analysis of the \bb -- decay amplitudes by
the proposed method can be the evaluation of amplitude (9) with the use
of the intermediate-state energies and wave functions calculated within
the QRPA. As mentioned above, the main reason for the hindrance of the
\bb -- decay amplitude, which is caused by
existence of the GTS, has been taken into account in eq. (9). That is
why we do not expect of essential change of amplitudes $M_{GT}$ as
compared with calculated ones. A consideration for the particle-hole
interaction (for instance, in form (\ref{LM})) will cause to some
redistribution of the small GT strength over the low-lying intermediate
states (AGTS) without the change of their total strength. This latter
is mainly determined by the difference of the sum rule $3e_q^2(N-Z)$
and the strength exhausted by the GTS.  Nevertheless, because the \bb
-- decay amplitude is formed at the expence of a rather small number of
the AGTS (this statement follows from the analysis of numerical
calculations of $M_{GT}$), the consideration for the quasiparticle
interaction can result in some change of $M_{GT}$. The further analysis
will allow to specify these reasons and, probably, to improve the
description of experimental data.

5. The authors are grateful to S.A.Fayans for discussions.

This work is supported in part by Grant No.MQ2300 from the
International Science Foundation and Russian Government, by Grants
Nos.95-02-05917a and 96-02-17596 from Russian Foundation for Basic
Researches. One of the authors (M.H.U.) is grateful to the
International Soros Science Education Program for support (Grant 444p
from the Open Society Institute, New York).

\newpage
\textwidth=19cm
\oddsidemargin=-1.5cm
Table 1. Explanations are given in the text.
\medskip

\begin{tabular}{|c|c|cc|c|c|c|cc|c|}
\hline
parent & type of & $G_{2\nu}$ & & $M_{GT}^{exp.}$ & $M_{GT}$ & $h_{GT}$
& $T_{1/2}^{exp.}$ & & $T_{1/2}^{calc.}$ \\ nucleus & decay &
$years^{-1}\ m_e^2$ & & $m_e^{-1}$ & $m_e^{-1}$ & & $years$ & & $years$
\\
\hline
$^{76}Ge $ & $\beta^-\beta^-$ & $1.317\times 10^{-19}$ & \cite{Doi88} &
0.0919 & 0.0387 & 0.131 & $0.9 \times 10^{21}$ & \cite{Vas90} & $5.0
\times 10^{21}$ \\
& & & & 0.0737 & & & $1.43\times 10^{21}$ & \cite{Bal92} & \\ $^{78}Kr
$ & $ec\ ec$ & $1.957\times 10^{-21}$ & \cite{Doi92} & & 0.0285 & 0.090
& & & $6.2 \times 10^{23}$ \\ & $\beta^+ec$ & $1.174\times 10^{-21}$ &
\cite{Doi92} &                  &                   &         &                      &              & $1.0 \times 10^{24}$ \\
$^{82}Se $ & $\beta^-\beta^-$ & $4.393\times 10^{-18}$ & \cite{Doi88} &
0.0459 & 0.0295 & 0.093 & $1.08\times 10^{20}$ & \cite{Ell92} & $2.6
\times 10^{20}$ \\
$^{96}Zr $ & $\beta^-\beta^-$ & $1.953\times 10^{-17}$ & & 0.0362 &
0.0678 & 0.324 & $3.9 \times 10^{19}$ & & $1.1 \times 10^{19}$ \\
$^{96}Ru $ & $ec\ ec$ & $6.936\times 10^{-21}$ & \cite{Doi92} & &
0.1005 & 0.338 & & & $1.4 \times 10^{22}$ \\ & $\beta^+ec$ &
$1.148\times 10^{-21}$ & \cite{Doi92} & & & & $>6.7\times 10^{16}$ &
\cite{Nor84} & $8.6 \times 10^{22}$ \\
$^{100}Mo$ & $\beta^-\beta^-$ & $9.553\times 10^{-18}$ & \cite{Doi88} &
0.0954 & 0.1606 & 0.329 & $1.15\times 10^{19}$ & \cite{Eji91} & $4.1
\times 10^{18}$ \\
$^{106}Cd$ & $ec\ ec$ & $1.573\times 10^{-20}$ & \cite{Doi92} & &
0.1947 & 0.319 & & & $1.7 \times 10^{21}$ \\ & $\beta^+ec$ &
$1.970\times 10^{-21}$ & \cite{Doi92} & & & & $>6.6\times 10^{18}$ &
\cite{Barab} & $1.3 \times 10^{22}$ \\
$^{116}Cd$ & $\beta^-\beta^-$ & $8.000\times 10^{-18}$ & \cite{Mor94} &
0.0754 & 0.0788 & 0.258 & $2.25\times 10^{19}$ & \cite{Mor94} & $1.2
\times 10^{19}$ \\
$^{124}Xe$ & $ec\ ec$ & $5.101\times 10^{-20}$ & \cite{Doi92} & &
0.0528 & 0.085 & & & $7.0 \times 10^{21}$ \\ & $\beta^+ec$ &
$4.353\times 10^{-21}$ & \cite{Doi92} & & & & & & $8.2 \times 10^{22}$
\\
$^{128}Te$ & $\beta^-\beta^-$ & $8.624\times 10^{-22}$ & \cite{Doi88} &
0.0123 & 0.0529 & 0.090 & $7.7 \times 10^{24}$ & \cite{Ber93} & $4.1
\times 10^{23}$ \\
$^{130}Te$ & $\beta^-\beta^-$ & $4.849\times 10^{-18}$ & \cite{Doi88} &
0.0087 & 0.0468 & 0.085 & $2.7 \times 10^{21}$ & \cite{Ber93} & $9.4
\times 10^{19}$ \\
$^{130}Ba$ & $ec\ ec$ & $4.134\times 10^{-20}$ & \cite{Doi92} & &
0.0568 & 0.082 & $>4 \times 10^{21}$ & \cite{Bar95} & $7.5 \times
10^{21}$ \\ & $\beta^+ec$ & $1.387\times 10^{-21}$ & \cite{Doi92} & & &
& $>4 \times 10^{21}$ & \cite{Bar95} & $2.2 \times 10^{23}$ \\
$^{136}Xe$ & $\beta^-\beta^-$ & $4.870\times 10^{-18}$ & \cite{Doi88} &
0.0299 & 0.0341 & 0.088 & $>2.3\times 10^{20}$ & \cite{Vui93} & $1.1
\times 10^{20}$ \\
$^{136}Ce$ & $ec\ ec$ & $3.988\times 10^{-20}$ & \cite{Doi92} & &
0.0512 & 0.081 & & & $9.6 \times 10^{21}$ \\ & $\beta^+ec$ &
$6.399\times 10^{-22}$ & \cite{Doi92} & & & & & & $6.0 \times 10^{23}$
\\
$^{150}Nd$ & $\beta^-\beta^-$ & $1.200\times 10^{-16}$ & \cite{Doi88} &
0.0221 & 0.0642 & 0.182 & $1.7 \times 10^{19}$ & \cite{Art93} & $2.0
\times 10^{18}$ \\
& & & & 0.0304 & & & $9 \times 10^{18}$ & \cite{Moe92} & \\
\hline
\end{tabular}

\end{document}